\documentclass[journal]{IEEEtran}

\usepackage{lineno,hyperref,multirow,graphicx,booktabs,array,ragged2e,subfig}
\usepackage[table]{xcolor}
\usepackage{dblfloatfix}
\usepackage{amssymb}
\usepackage{amsmath}
\usepackage{bm}

\modulolinenumbers[5]

\begin{document}

\expandafter\def\expandafter\UrlBreaks\expandafter{\UrlBreaks%  save the current one
  \do\a\do\b\do\c\do\d\do\e\do\f\do\g\do\h\do\i\do\j%
  \do\k\do\l\do\m\do\n\do\o\do\p\do\q\do\r\do\s\do\t%
  \do\u\do\v\do\w\do\x\do\y\do\z\do\A\do\B\do\C\do\D%
  \do\E\do\F\do\G\do\H\do\I\do\J\do\K\do\L\do\M\do\N%
  \do\O\do\P\do\Q\do\R\do\S\do\T\do\U\do\V\do\W\do\X%
  \do\Y\do\Z}
  
  \makeatletter
\newcommand*{\centerfloat}{%
  \parindent \z@
  \leftskip \z@ \@plus 1fil \@minus \textwidth
  \rightskip\leftskip
  \parfillskip \z@skip}
\makeatother

\newcolumntype{L}[1]{>{\RaggedRight\arraybackslash}p{#1}}
\newcolumntype{C}[1]{>{\centering\let\newline\\\arraybackslash\hspace{0pt}}m{#1}}
\newcolumntype{R}[1]{>{\RaggedLeft\arraybackslash}p{#1}}

\renewcommand{\tableautorefname}{Table}
\renewcommand{\figureautorefname}{Figure}
\renewcommand{\equationautorefname}{Equation}

\renewcommand{\sectionautorefname}{Section}
\renewcommand{\subsectionautorefname}{Section}
\renewcommand{\subsubsectionautorefname}{Section}
\renewcommand{\tablename}{Table}
\renewcommand{\figurename}{Figure}
\newcommand{\subfigureautorefname}{\figureautorefname}

\title{An Empirical Study into the Success of Listed Smart Contracts in Ethereum}

\author{
    \IEEEauthorblockN{Pieter Hartel, Ivan Homoliak, Dani\"el Reijsbergen} \\[0.1cm]
    \IEEEauthorblockA{Singapore University of Technology and Design}
}

\maketitle

\begin{abstract}

	Since it takes time and effort to put a new product or service on the market, one would like to predict whether it will be a success.
	In general this is not possible, but it {\it is} possible to follow best practices in order to maximize the chance of success.
	A smart contract is intended to encode business logic and is therefore at the heart of every new business on the Ethereum blockchain.
	We have investigated how to measure the success of smart contracts, and whether successful smart contracts have characteristics that less successful smart contracts lack.
	The appearance of a smart contract on a listing website such as Etherscan or StateoftheDapps is such a characteristic.
	
	In this paper, we present a three-pronged analysis of the relative success of listed smart contracts.
	First, we have used statistical analysis on the publicly visible transaction history of the Ethereum blockchain to determine that listed contracts are significantly more successful than their unlisted counterparts.
	Next, we have conducted a survey among more than 200 developers via an anonymous online survey about their experience with the listing process.
		A significant majority of respondents do not believe that listing a contract itself contributes to its success, but they believe that the extra attention that is typically paid in tandem with the listing process {\it does} contribute. 
	Finally, based on the respondents' answers, we have drafted 10 recommendations for developers and validated them by submitting them to an international panel of experts.
\end{abstract}

\begin{IEEEkeywords}
~
Subject categories: Computers and information processing, Social implications of technology \\[1ex]
Additional manuscript keywords:  Blockchain, Ethereum, Smart contracts, New product development, Business success, Recommendations for developers
\end{IEEEkeywords}

%\linenumbers

\section{Introduction}
\label{sec:introduction}
A \emph{Blockchain} is a peer-to-peer database that contains the full history of the database transactions, and whose contents is cryptographically secured. A \emph{Smart Contract} is a program that runs on such a database. In particular, the contract creation, and all subsequent interactions with the contract --- i.e., function calls --- are stored on the blockchain. Every node in the peer-to-peer network can create contracts and send method calls at a certain \emph{cost} in terms of the blockchain's native token. Each node also executes every method call in the blockchain's history, and hence maintains an accurate and shared view of the global state.
Smart contracts typically codify the \emph{business logic} of a blockchain application. For example, a lottery contract contains logic that decides when the player wins the jackpot, and what percentage of each bet will go to the owner of the lottery. There are hundreds of blockchains, but here we focus on Ethereum~\cite{buterin2014next}, which is the most prominent blockchain that offers smart contracts. There are many thousands of Ethereum developers writing new applications for a variety of domains (See \autoref{tab:ethereum} for an up-to-date breakdown of domains). Henceforth we will use the term ``developers'' in a broad sense to include engineers, managers, owners, business developers, marketers, etc.

Millions of smart contracts have already been deployed on Ethereum.
Consistent with all human endeavour, the developers of smart contracts are striving for success. But what can developers do to foster the success of their smart contracts? What skills and resources do they need? To address these questions, we will analyse both historic data from the blockchain, and survey data obtained from a questionnaire sent to developers. Based on our analysis, we will provide validated recommendations for the Ethereum community. 

\paragraph*{New Product Development} To avoid common pitfalls when developing a questionnaire, it is helpful to follow a well-established social research framework. In this paper, we study the success of smart contracts through the lens of New Product Development (hereafter NPD). The objective of NPD is to {\it manage} the process of new product development, rather than to have to depend on intuition, flair and luck. NPD has been researched intensively for products~\cite{Cooper1987}, services~\cite{Johne1998}, internet-based services~\cite{Menor2002}, mobile apps~\cite{Datta2013}, but not yet for smart contracts. 

Products and services differ in some respects. For example, services are intangible, and more heterogeneous than products~\cite{Menor2002}. However, the most relevant difference for our research is the time scale on which development takes place. Product development can take years, while sometimes only a few months are available for developing an internet-based service~\cite{Menor2002}. This is because the barrier to market entry for services is lower than for products. This barrier is even lower for smart contracts, because anyone who can program can also download the open source tools, such as the Truffle framework\footnote{\url{https://www.trufflesuite.com/}}, and develop and deploy a smart contract. Currently, an estimated 10,000 new smart contracts are deployed every day on Ethereum\footnote{\url{https://console.cloud.google.com/bigquery?p=bigquery-public-data&d=ethereum_blockchain&t=contracts&page=table}} alone. About the same number of mobile apps are launched globally every day~\cite{Koch2017}. However, the app economy~\cite{Mendelson2018} is many times larger than that of Ethereum\footnote{\url{https://etherscan.io/stat/supply}}. Therefore we posit that the pressure on rapid deployment of smart contracts is even higher than on mobile apps.

Products and services also have a lot in common. For example, developers can take specific {\it measures} to promote success. The most important measure is to ensure {\it product advantage}, which means that the product or service has an advantage over its competition --- e.g., through higher quality, lower cost, or innovation. There are other measures, such as technological and market synergies that describe to what extent the developers have the skills and resources needed for the development, production, marketing, etc. NPD also provides a number of {\it metrics} of success, such as market share and payback period. Again we will focus on the most important ones, which are market share, profitability, and sales~\cite{Cooper1987}. We discuss later how these metrics are best observed in the context of historic blockchain data.

\paragraph*{Blockchain analysis} A blockchain is an attractive research object because the full history of all contracts is publicly available. For each smart contract we can query which transactions have ever taken place. We can also query for each transaction who was the sender and who was the recipient, when the transaction took place, and what the value of the transaction was. This provides the data needed to calculate NPD success metrics. Additionally, there are various listing websites, such as Etherscan\footnote{\url{https://etherscan.io/contractsVerified}}, that allow developers to \emph{register} smart contracts --- registered smart contracts are then {\it listed} on the site. Given a selection of listing sites, we can determine for each contract when and where it has been listed. Listing is usually undertaken as part of a marketing campaign. We assume that developers do not invest time and resources into a marketing campaign for an inferior product, but that such a campaign is part of a broader package to achieve product advantage.

This brings us to the technical core of our approach. We will collect pairs of contracts that resemble each other as much as possible in terms of typical code similarity metrics, but that otherwise have one major difference: one of the pair is listed and the other is unlisted. One prominent reason between this difference would be that the developers of the listed contract believed that it had sufficient product advantage to justify the effort needed to list the contract, whereas the developers of the unlisted contract did not. NPD theory then predicts that the listed contract should be more successful than the unlisted contract. Indeed, we have found that listed contracts on the Ethereum blockchain are considerably more successful than unlisted contracts. 

\paragraph*{Questionnaire analysis} To investigate the role of the listing process in achieving business success, we have distributed an anonymous questionnaire to Ethereum developers with more than 200 completed forms as a result. Given the high pressure to put something on the market quickly, we wanted to be sure that developers had no trouble with the listing process. We have therefore used the theory of the Technology Acceptance Model (hereafter TAM)~\cite{Venkatesh2000} to investigate the usefulness and perceived ease of use of the listing process. We also wanted to know which measures developers think are contributing to the success of a smart contract. Hence, we asked developers whether they think that listing a smart contract contributes to its success, and more generally whether devoting special attention to a contract does so.

\paragraph*{Contributions} We make the following contributions:
\begin{itemize}
\item Our study is the first to investigate the applicability of the theories of NPD and TAM to smart contract development.
\item We compare the success of a series of random samples of listed and unlisted smart contracts. 
\item We conduct a survey amongst 200+ Ethereum developers, investigating their opinion on the contribution of listings services to the success of smart contracts. More generally our survey provides new insights in to the state of the Ethereum eco-system.  
\item We provide a set of validated recommendations for Ethereum developers to increase the chance of success of the technology and its applications.
\end{itemize}

The paper is structured as follows. In \autoref{sec:background} we discuss the background of our work and present our research questions. In \autoref{sec:method} we present the data collection methodology for the blockchain data and the questionnaire, and \autoref{sec:results} presents the results. Based on our analysis of the literature and the results, we list a number of validated recommendations in \autoref{sec:recommendations}. \autoref{sec:limitations} lists the threats to validity of our work, and the last section concludes the paper.

\section{Background \& Preliminaries}
\label{sec:background}
Developers of smart contracts can try to achieve product advantage by using technical skills and resources to analyse and improve the functionality, usability, and security of their contracts. For example developers can thoroughly test a contract, they can hire an independent auditor to perform a code audit, or they can upload the source code of a contract to a website such as ChainSecurity\footnote{\url{https://securify.chainsecurity.com}}, which checks the contract for a range of known security vulnerabilities.

Bosu et al.~\cite{Bosu2019} report on a detailed study of blockchain related software engineering practices. In December 2017 and January 2018 a survey was held among 156 active developers of blockchain software. The demographics of the Bosu study has strong similarities with ours. The response rate is also about the same, and the respondents are just as well trained as ours. Bosu's respondents have slightly more software development experience but slightly less experience in developing blockchain-specific software. The latter may be because the data from the Bosu study is more than a year older than our data. We interpret the agreement in demographics as an indication that our data is as representative of the eco-system as the Bosu data.

The Bosu study and ours are complementary in the sense that we focus on listing, which is a marketing instrument, while Bosu et al.\ focus on the actual software development aspects. One of the most important conclusions from the work of Bosu et al.\ is that the tooling and documentation across the board of blockchain technologies is immature. We hope to learn from our study whether more than a year later there is at least some increase in maturity of Ethereum tooling.

There are several sites where smart contracts can be listed, the most important of which is Etherscan. To list a contract, the developer must complete a form on the Etherscan website, and upload the source code of the contract to the website. Etherscan then checks that the source code and the byte code deployed on the block chain are consistent. If the byte code produced by the Solidity compiler is identical to the byte code already deployed on the Ethereum block chain, the contract is considered ``verified'', and it is then listed on the Etherscan website. The business logic of a listed smart contract is available for public scrutiny, which should give a listed contract an advantage over competing contracts that have not been listed.

The status of listing a contract is shown on the Etherscan website with a trust seal:
\begin{figure}[!ht]
\vskip-0.3cm
\includegraphics[width=0.45\textwidth]{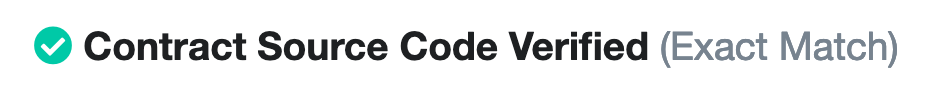}
\end{figure}
\vskip-0.3cm
Prior research has found that trust seals have limited impact on the behaviour of visitors to websites \cite{Kirlappos2012}, because end-users tend not to notice the trust seals, and even if they do, they do not necessarily understand what it means. Therefore we focus not on the end-user but on the developers of smart contracts. Based on NPD theory, we hypothesize that developers take a range of measures to achieve product advantage, and that listing a contract on Etherscan (or other, similar websites) helps to reach product advantage.

Developers who want to turn their smart contracts into a success must be well-trained and have the right experience. Because blockchain technology is relatively young, developers often have little experience, and the technology is not yet mature. We discuss a number of papers that have raised these issues.

Based on an analysis of all 11M+ external and internal smart contracts on Ethereum prior to December 16, 2018, Di Angelo and Salzer~\cite{Angelo2019} state that the differences between smart contracts and regular programs pose problems for code reuse. For example, a smart contract has a different life cycle than a normal program because a smart contract can no longer be changed once it is deployed on the blockchain. Developers therefore often use a contract with a fixed address as a gatekeeper, which redirects all calls to the current version of the contract in question. This is inconvenient and can cause problems. The Ethereum library mechanism is also relatively rigid, so that much more code is copied than would be necessary with a more flexible, parameterized mechanism.

Pinna et al.~\cite{Pinna2019,Tonelli2018} analyse all 12K+ verified smart contracts on Etherscan prior to January 24, 2018. Here too, the differences between smart contracts and regular programs challenge the developer. For example, developers are more or less forced to regularly switch to a new version of the compiler to be able to benefit from bug fixes. Standard code metrics (e.g. source lines of code) generally have low values, but high variance; this is a sign that many smart contracts are just prototypes, and that the developer community is not homogeneous. Pina et al\. also think that the most commonly used smart contracts are not necessarily the best-written contracts.

On January 1, 2017, the number of verified smart contracts on Etherscan was still so small ($N=811$) that Bartoletti and Pompianu~\cite{Bartoletti2017a} were able to classify all contracts manually. The majority of contracts had a financial purpose, and the second biggest category was entertainment. Within each category there are standard patterns according to which most contracts are made, such as the ERC20 token contract. It is conducive to the quality of the work if a developer can use such a design pattern.

Since 2017 a number of papers have been published~\cite{Porru2017,Destefanis2018} with proposals for a new discipline called Blockchain Oriented Software Engineering (BOSE). The idea is that as many best practices as possible are collected and synthesized into manageable patterns and knowledge\footnote{\url{https://www.computer.org/education/bodies-of-knowledge/software-engineering}}.

\paragraph*{Smart contracts, tokens or DApps} A token is a special type of smart contract that represents intrinsic value to the organisation that issues the token. A DApp is a Distributed Web Application built on top of a smart contract so that end-users can interact with the contract. The key component is always the smart contract; hence we will usually talk about a smart contract even in a context where the more common term would be a token or DApp. We will also use the word smart contract when discussing a DApp listing service, or a Token listing service.

\paragraph*{Listing} We will henceforth talk about listed smart contracts when a contract has been listed on Etherscan, a DApp listing service, or a Token listing service. Our use of the word listing should not be confused with the legal term ``listed security'', which may even apply to crypto currencies, such as Ether, Ethereum's native token. The listed contracts are available from Etherscan, the DApps are available from StateoftheDapps\footnote{\url{https://www.stateofthedapps.com/rankings/platform/ethereum}}, and the tokens are available from ForkDelta\footnote{\url{https://github.com/forkdelta/tokenbase}}.

\subsection{Research questions} \label{sec:research_questions}
According to NPD theory, developers of listed smart contracts are investing in marketing to amplify their product advantage. In contrast to developers of unlisted contracts, they are happy to go through a listing process. However, we do not know to what extent this extra effort pays off. The research questions that we will try to answer are therefore:
\begin{enumerate}
\item To what extent are listed smart contracts more successful than unlisted smart contracts? In particular, we investigate the following hypotheses:
	\begin{enumerate}
	\item Listed smart contracts have a \emph{higher number of transactions} than unlisted smart contracts.
	\item Listed smart contracts receive \emph{more ether} than unlisted smart contracts.
	\item Listed smart contracts have a \emph{higher number of distinct users} than unlisted smart contracts.
	\end{enumerate}
\item According to developers, which measures contribute most to success? We investigate the following hypotheses: that a majority of developers
	\begin{enumerate}
	\item think that listing services are \emph{helpful}.
	\item think listing services \emph{easy to use}.
	\item think that listed contracts are intrinsically \emph{better}.
	\item think that listed contracts are \emph{more successful}.
	\end{enumerate}
\end{enumerate}
We will answer the first research question by analysing and comparing listed and unlisted smart contracts on the Ethereum blockchain. Then we will answer the second  research question via a survey of Ethereum developers. Finally, the results of the analysis will be synthesised into a set of recommendations for developers.

\section{Method}
\label{sec:method}
We discuss the methods used to answer the two research questions.

\begin{figure}[!ht]
\centerfloat
\framebox[0.50\textwidth]{
\begin{minipage}[t]{0.48\textwidth}
\begin{flushleft}
 SELECT \\
 $\;\;\;\;\;\;$ $\ast$ \\
 FROM \\
 $\;\;\;\;\;\;$ `bigquery-public-data.ethereum\_blockchain.traces' AS contracts \\
 WHERE $ $ contracts.trace\_type = `create' \\
 $\;\;\;\;\;\;$ AND contracts.status = 1 \\
 $\;\;\;\;\;\;$ AND contracts.to\_address IS NOT NULL \\
 $\;\;\;\;\;\;$ AND DATE(contracts.block\_timestamp) $<$ \\
 $\;\;\;\;\;\;\;\;\;\;\;\;$ `2019-01-01';

\end{flushleft}
\end{minipage}
}
\caption{Google BigQuery statement to discover all internal and external smart contracts on Ethereum that were deployed before 1 Jan 2019.}
\label{fig:bigquery}
\end{figure}

\begin{table*}[!hb]
	\caption{Typical differences in the source text of a matched pair.}
	\centerfloat
	\begin{tabular}{c|p{0.4\textwidth}|p{0.4\textwidth}}
		\parbox[t]{2mm}{\rotatebox[origin=c]{90}{URL$\;\;\;$}}
			& \footnotesize \url{https://etherscan.io/address/0xf5b9fcc3e47cbbd65112f392056b1e124a2747ac#code}
									& \footnotesize \url{https://etherscan.io/address/0xd2680e5287dfa6T7cb3eb279ed5752ed593153fea#code } \\ \hline
		\parbox[t]{2mm}{\multirow{6}{*}{\rotatebox[origin=c]{90}{Text diffs}}}
			& \footnotesize contract ShareCoin\{		& \footnotesize contract PlayCoin\{ \\ \cline{2-3}
			& \footnotesize function ShareCoin() \{		& \footnotesize function PlayCoin()\{ \\
			& \footnotesize $\;$ balanceOf[msg.sender] = 60000000000000;
									& \footnotesize $\;$ balanceOf[msg.sender] = 1280000000000; \\
        		& \footnotesize $\;$ totalSupply = 60000000000000;
									& \footnotesize $\;$ totalSupply = 1280000000000; \\
        		& \footnotesize $\;$ name = ``Share Coin'';	& \footnotesize $\;$ name = ``Play Coin''; \\
        		& \footnotesize $\;$ symbol = "SC";		& \footnotesize $\;$ symbol = ``PLAYC''; \\
	\end{tabular}
\label{tab:matched_pair}
\end{table*}

\setlength{\tabcolsep}{1pt}
\begin{table}[!h]
\caption{Contingency table of the results of the heuristic step versus the expensive NLD step}
\centerfloat
\begin{tabular}{L{0.115\textwidth}L{0.13\textwidth}L{0.11\textwidth}L{0.125\textwidth}}
				& The number of basic blocks is different
								& The number of basic blocks is similar
												& Total \\
\toprule
				The NLD of the & \textit{true negatives}	& \textit{false positives} \\
opcodes is large & \textbf{98.5\%} (N=114633)	& \textbf{1.5\%} (N=1736)	& \textbf{100\%} (N=116369) \\ 
\midrule
				The NLD of the & \textit{false negatives}	& \textit{true positives} \\
opcodes is small & \textbf{17.4\%} (N=90)	& \textbf{82.6\%} (N=427)	& \textbf{100\%} (N=517) \\
\midrule
Total				& \textbf{98.1\%} (N=114723)	& \textbf{1.9\%} (N=2163)	& \textbf{100\%} (N=116886) \\
\end{tabular}
\label{tab:nld_cross_table}
\end{table}
\setlength{\tabcolsep}{6pt}

\subsection{Blockchain analysis}
Google's BigQuery service\footnote{\url{https://cloud.google.com/bigquery}} provides a convenient but relatively expensive interface to the live Ethereum blockchain via SQL queries such as the one in \autoref{fig:bigquery}.

The same SQL query but with the constraint on \texttt{trace\_type} omitted will yield all transactions ever executed on Ethereum before the given date. To collect relevant data, such as the name and the registration date on listed smart contracts we scraped the websites Etherscan, StateoftheDapps and ForkDelta and stored the data in Google BigQuery tables. By a SQL join operation of these scraped tables on the public \texttt{ethereum\_blockchain.traces} table from BigQuery, all transactions sent to listed and unlisted contracts are available for analysis with an SQL query.

\paragraph*{Functional similarity at source code level} A comparison of the success of listed and unlisted contracts must be fair. To achieve this we need a method to find an unlisted contract that is functionally similar to a given listed contract.  This is the same problem as detecting contract clones, for which a range of algorithms has been proposed by researchers and anti-virus companies~\cite{Bellon2007}. The example in \autoref{tab:matched_pair} shows the properties of such a matched pair. The links provided point at the source code of the contracts on Etherscan. The differences indicated in the table are indeed small: the name of the contract and the values of the parameters have been changed but everything else is the same. 

\paragraph*{Edit distance} An edit distance, such as the Levenshtein distance of the source code of two contracts is generally considered to be a good similarity measure~\cite{He2019}. The edit distance is the number of character insertions and deletions that have to be made to one text to change it into another. Normalisation then divides the distance by the maximum number of characters of the two texts. The Normalised Levenshtein Distance (NLD) of the two contracts of \autoref{tab:matched_pair} is 0.0083, which means that less than 1\% of the source text of the two contracts is different.

\paragraph*{Functional similarity at opcode level} Since the source code of unlisted smart contracts is not necessarily publicly available, but the bytecode is always available, we will use a variant of bytecode similarity as a proxy for source code similarity. As shown in \autoref{tab:matched_pair}, most of the expected differences of similar contracts are in the data. Therefore, like di Angelo and Salzer~\cite{Angelo2019}, we ignore the arguments of the EVM PUSH instruction and use only the opcodes in the similarity calculations.

\paragraph*{Algorithm for finding matched pairs} Since the NLD calculation is relatively expensive and Google BigQuery is not a free service, we have devised a heuristic to filter contracts that are unlikely to be similar. The proposed algorithm with the heuristic for finding matched pairs is as follows:
\begin{enumerate}
\item	Given a contract $C$ that was mined in a certain month $M$, with a bytecode $B$, with $N$ basic blocks, and with a list of opcodes $O$.
\item	Find all contracts $C'$ with $M'=M$, $B' \neq B$, and $|N'-N| \leq 1$. {\bf (Heuristic.)}
\item	If $NLD(O,O') \leq 0.1$ we have found a matched pair $(C,C')$.\end{enumerate}
Step 2 is the inexpensive heuristic that retains only pairs with an almost similar number of basic blocks. Step 3 is the expensive step that computes the NLD to discard false positives retained by the heuristic. To show that the heuristic is effective, we have taken a uniform random sample of 484 verified smart contracts from all 1485 verified smart contracts mined in a randomly selected month (October 2017). We have calculated the NLD on the source code pairs and on the opcode pairs and of all $484 \times 483/2=116886$ different pairs of smart contracts from the sample.

\paragraph*{Efficiency trade-off} \autoref{tab:nld_cross_table} shows a contingency table of the number of contract pairs in the four relevant conditions defined by the heuristic and NLD selections. The second column lists the number of pairs that have been discarded by the heuristic. Only 90 (17.4\%) pairs were false negatives and discarded incorrectly. Therefore 17.4\% of the matching pairs will not found with our heuristic. Also 1736 (1.5\%) potential matching pairs are false positives because they are retained by the heuristic, and must be discarded by the expensive NLD step. The heuristic therefore reduces the amount of computation time needed by 98.5\%, but it misses 17.5\% of the matched pairs. We found this the most useful trade-off between saving time (and expenses) and loosing matched pairs. He et al.~\cite{He2019} use a different trade-off, which, when applied to the same data, has no false negatives but 28.7\% false positives.

\paragraph*{Validation of opcode similarity} Even with a low NLD of the opcodes, a matched pair could still be a false positive when the source codes are significantly different. Therefore, to verify that opcode similarity as defined above is an appropriate proxy for source code similarity, we manually compared the sources of each of the 427 true positive opcode-similar pairs to inspect the differences. The following differences occur frequently:
\begin{itemize}
\item	Name and parameters of the contract are changed.
\item	The \texttt{SafeMath} contract is added to one of the contracts, but not actually used. Therefore the opcodes do not change.
\item	The constructor function is sometimes given more arguments to increase flexibility.
\item	Additional events are occasionally emitted to improve the communication between the contract and the DApp.
\end{itemize}
All of these changes maintain the structure and the functionality of the contract. This confirms that opcode similarity is a good proxy for source code similarity. 

\paragraph*{Proxies for the NPD success metrics} We will use technical proxies for the NPD success metrics that can be extracted from the block chain. For each proxy we also discuss potential bias.
\begin{itemize}
\item	For sales we use the \emph{number of transactions} sent to a contract. This may underestimate sales as the contract may receive transactions on additional addresses that it manages.
\item	For market share we use the \emph{number of unique interacting addresses} from which the transactions are sent. This may overestimate the market share, as one person may own different address. 
\item	For profits we use the \emph{total amount of Ether} sent to a contract. This may underestimate the profit as the contract may manage additional addresses that receive ether as well.
\end{itemize}
Other success measures, such as payback period~\cite{Cooper1987} cannot be estimated with technical proxies, so our assessment of success is necessarily partial, and approximate because of the bias in the proxies.

\paragraph*{Data collection} To assess the success of a smart contract as measured by our proxies we take a random sample of contracts listed and mined in eight months: Jan, Apr, Jul and Oct of the years 2017, and 2018. We analysed 8 out of 48 months of Ethereum blockchain data, and took a random sample for each of the 8 months, rather than all contracts of a particular month, because BigQuery is not a free service. Most contracts are listed shortly after they are mined, so the requirement that a contract is listed and mined in the same month is not a severe restriction. For each contract, we collected all transactions sent to the contract in the 100 days~\cite{Sigg2019} after the contract was mined. We expect most contracts to generate a lot of business when they are new. Some contracts such as Initial Coin Offerings (ICO) have a limited lifetime, so there should be more transactions early on during the lifetime of the contract. We paired each listed contract with a similar unlisted contract mined in the same month as the listed contract and also collected all transactions of the unlisted contract for 100 days from the moment the unlisted contract was mined. Then for each pair we compared the success metrics of the listed and unlisted contracts. 

\subsection{Questionnaire analysis}

\paragraph*{Technology acceptance model} Listing services are a technology and as such it makes sense to investigate the acceptance of this technology by developers using the Technology Acceptance Model (hereafter TAM)~\cite{Venkatesh2000}. The questionnaire contains 6 items on ``Perceived usefulness'' and 6 items on ``Ease of use''. We use the standard items from the TAM model but tailored them to the technology of interest. For example, ``I find a listing service useful in my job as a developer'' and  ``Learning to operate a listing service is easy for me''. Each item is scored on a 5-point Likert scale (Strongly disagree, Disagree, Neutral, Agree, Strongly agree).

\paragraph*{New Product Development} 
In the NPD literature it is not unusual to construct a large model to explore the relationship between the measures that promote success and the metrics for success~\cite{Cooper1987,Atuahene1996}. The disadvantage of using comprehensive models is that the questionnaires tend to be long. In order to keep the survey relatively short (23 questions), we have chosen to ask the opinion of respondents about the relationship between measures that developers can take and success metrics. We propose two constructs as follows.

\textit{Listing improves success.} This construct consists of 9 items (scored on the same 5-point Likert scale as before):

\begin{quote}
To what extent do you agree with the following statements about listed smart contracts, DApps and/or tokens on Etherscan?
	\begin{enumerate}
	\item	They generate transactions more quickly than unlisted ones.
	\item	They generate more transactions than unlisted ones.
	\item	They generate more revenue than unlisted ones.
	\item	They have more customers than unlisted ones.
	\item	They are more reliable than unlisted ones.
	\item	They are more efficient than unlisted ones.
	\item	They are more secure than unlisted ones.
	\item	They are more maintainable than unlisted ones.
	\item	They are more successful than unlisted ones.
	\end{enumerate}
\end{quote}

Items (1-4) incorporate the NPD success metrics that we focus on. Items (5-8) represent the NPD measure product advantage, and specifically that the product must be of higher quality than competing products~\cite{Cooper1987}. Because a smart contract is software, we assume here that quality \textit{is} software quality~\cite{Kitchenham1996}. Item 9 is a control question to check consistency of the responses. The overall score for ``Listing improves success'' is the average of the 9 items.

\textit{Attention improves success.} Listing is part of a broader package of technical and business related measures that developers can take to achieve product advantage. Such activities tend to take more time than listing a contract and one would only undertake this if the contract merits this extra effort to achieve product advantage. Therefore we define a second construct ``Attention improves success'' with 4 items (on the same 5-point Likert scale as before):

\begin{quote}
	To what extent do you agree with the following statements about smart contracts, DApps and/or tokens?
	\begin{enumerate}
	\item	They are more successful if they have been formally verified.	
	\item	They are mode successful if they have passed a security audit.	
	\item	They are more successful if they have been open sourced.	
	\item	They are more successful if they have been listed.
	\end{enumerate}
\end{quote}

Items (1-3) relate success to a specific software engineering activity that we assume contributes to product advantage. There is also an open question at the end of the survey in which we give respondents the opportunity to identify other measures that they undertake to promote the success of their work. Again the last item is a control question to see if the respondents answer somewhat consistently. 

\paragraph*{Data collection} We sent our questionnaire to Ethereum developers during the months of April and May of 2019. First, StateoftheDapps sent about 1,000 potential respondents an email, but only a few completed the questionnaire. Then we queried LinkedIn for members with Ethereum and Solidity listed on their profile\footnote{\url{https://www.linkedin.com/search/results/people/?keywords=ethereum solidity}}. This query yielded about 11K LinkedIn members in March 2019. The first 2K of those have been asked individually to connect to one of the authors. 932 LinkedIn members accepted the connection request. These connections were then asked via a personal message to complete the questionnaire. In total we received 376 responses, with an overall response rate of about 12.5\%.

\paragraph*{Ethical considerations} The Institutional Review Board of our University has granted approval for the research under number IRB-19-00205.

\section{Results}
\label{sec:results}
We present the results for the two research questions.

\subsection{Blockchain analysis}
\begin{figure}[!ht]
\centering
\includegraphics[width=0.48\textwidth]{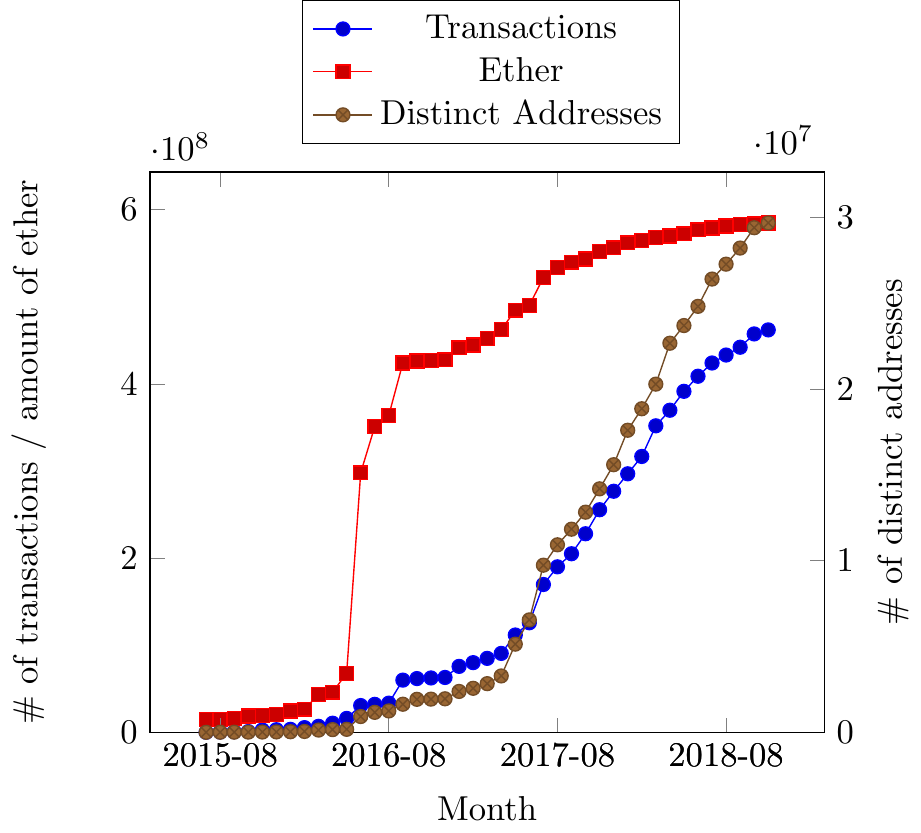}
\caption{Cumulative timelines of the three proxies until 1 Jan 2019. The attack on the DAO contract and the subsequent hard fork \cite{siegel2016understanding,Atzei2017} occurred in July 2016, which may explain the large amount of Ether sent in that month.}
\label{fig:time_cumulatives}
% \ih{This graph is very overloaded. There 3 different units in Y axis. And X axis also looks like it would have 2 different units. What about to split the graph? into 3 ones?}
% \pieter{These are all time lines on the same X-axis.}
% \pieter{I was wondering why the Ether graph is so different from the other two. What happened in July 2016 when 2.30184145983262E+08 Ether was transacted?}
% \daniel{July 2016 was the month of the DAO hard fork, which at least means that the 12 million ETH in the that contract had to be moved: \url{https://blog.ethereum.org/2016/07/20/hard-fork-completed/}, but that's still only 5\% f the total amount. 230 million ETH is quite a lot to be honest: even now, there's only about 107 million ETH total in circulation, and that's not even half (of course ETH can be moved several times in a single month.))}
\end{figure}

On 1 Jan 2019, there were 12,023,046 smart contracts on Ethereum with an average balance of 48.62 Ether. On average 38.42 transactions were sent to the contracts from on average 1.68 different addresses. \autoref{fig:time_cumulatives} shows cumulative time lines of the three proxies: ``nTx'' for the number of transactions sent to contracts, ``ether'' for the amount of ether sent to contracts and ``nDistinctFrom'' for the number of different addresses from which transactions were sent to contracts.

Also on 1 Jan 2019, 54,179 contracts were listed as listed on Etherscan, 5,187 smart contracts were listed on StateoftheDapps, and 1,017 Tokens were listed on ForkDelta. The overlap of Etherscan and StateoftheDapps consists of 1,451 smart contracts. Only 161 Tokens were neither listed on Etherscan nor on StateoftheDapps. We did not analyse the Tokens separately as they were almost all covered by the other listing methods.

\paragraph*{Listed contracts are more successful than unlisted contracts} \autoref{tab:proxies} shows for the months of October 2017 and October 2018 that about two orders of magnitude more transactions were sent to the listed contracts than to unlisted contracts, with a similar ratio for the other proxies. The difference between contracts listed on StateoftheDapps and those that are not listed is even more pronounced. We take this as evidence that, as determined by our proxies, listed contracts are orders of magnitude more successful than unlisted ones.

\begin{table*}[!ht]
	\centerfloat
	\begin{tabular}{L{0.25\textwidth}L{0.15\textwidth}L{0.15\textwidth}L{0.15\textwidth}L{0.15\textwidth}} \toprule
\bf October 2017			& \bf nTx / contract
							& \bf ether / contract
									& \bf nDistinct From / contract
											& \bf Total \# of contracts \\ \midrule
All contracts				& 8.1		& 7.5		& 0.6		& 424,233 \\
Listed contracts on Etherscan		& 1,214.0	& 1,072.1	& 105.7		& 1,372 \\
Unlisted contracts			& 4.2		& 4.0		& 0.3		& 422,861 \\
Listed contracts on StateoftheDapps	& 39,099.2	& 375.4		& 343.8		& 6 \\
Unlisted contracts			& 7.5		& 7.5		& 0.6		& 424,227 \\ \bottomrule \toprule
\bf October 2018			& \bf nTx / contract
							& \bf ether / contract
									& \bf nDistinct From / contract
											& \bf total \# of contracts \\ \midrule
All contracts				& 6.0		& 0.7		& 0.8		& 992,354 \\
Listed contracts on Etherscan		& 848.5		& 43.4		& 190.4		& 3,005 \\
Unlisted contracts			& 3.4		& 0.6		& 0.2		& 989,349 \\
Listed contracts on StateoftheDapps	& 667.9		& 91.3		& 42.3		& 253 \\
Unlisted contracts			& 5.8		& 0.7		& 0.7		& 992,101 \\ \bottomrule
	\end{tabular}
	\caption{Proxies for October 2017 and October 2018}
	\label{tab:proxies}
\end{table*}

\paragraph*{Detailed comparison of matched pairs}%
\phantomsection%
\label{detailed_comparison}%
To investigate the differences in success more deeply, we have taken a random sample of listed smart contracts and applied our algorithm to find a matching unlisted contract for every listed contract in the sample. To ensure that our samples are representative for the entire population, we compared key statistics of all 17602 listed smart contracts from 8 months (Jan, Apr, Jul, Oct for both 2017 and 2018) to all 812 listed smart contracts from our monthly samples. For both the samples and the population we found that (1) 80\% of all transactions are made by a small number of contracts, (2) 88\% of all contracts are registered on Etherscan on the same day as when the contract is deployed on the blockchain, and (3) more contracts are registered during working days than during the weekend. This confirms that the sample is representative.

Measured by the number of transactions, the listed contract of a pair is more successful in 83\% of the matched pairs, and the unlisted contract is more successful in 17\% of the pairs. We Googled the top 25 of these unlisted, successful contracts to check if they were listed on a site that we have not covered. Only 1 out of the 25 is a well-known Dapp.\footnote{\url{https://www.dapp.com/ko/edit_dapp/EtherCraft}}

\begin{figure*}[!ht]
\centerfloat
	\subfloat[][]{\includegraphics[width=0.3\textwidth]{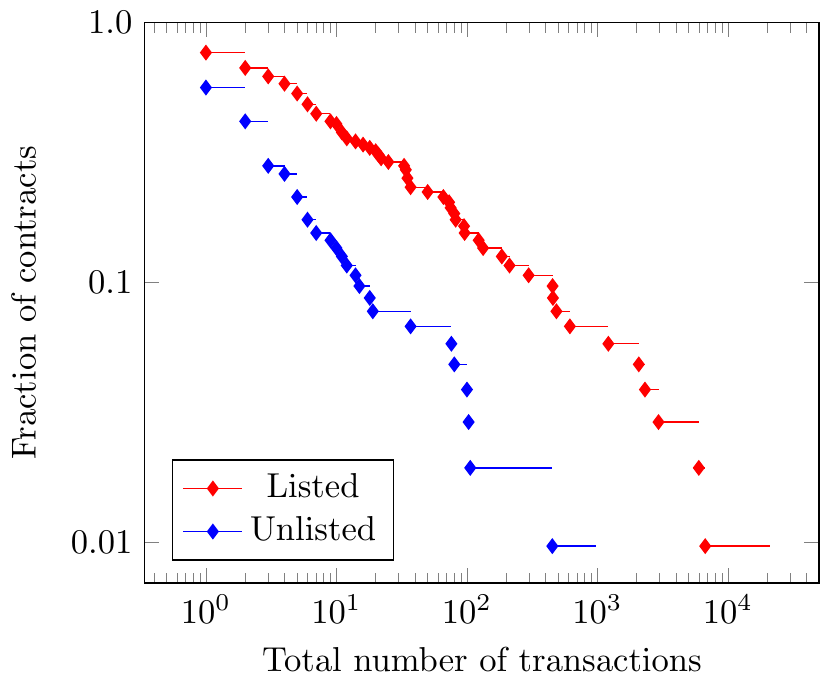}\label{fig:nTx_ecdf}}
	\hspace{0.02cm}
    \subfloat[][]{\includegraphics[width=0.3\textwidth]{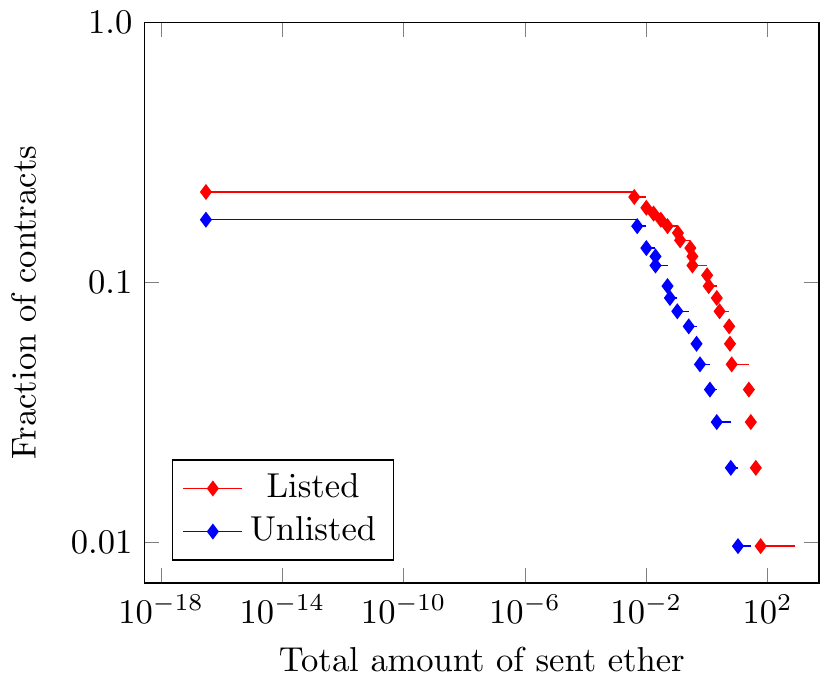}\label{fig:ether_ecdf}}
    \hspace{0.02cm}
    \subfloat[][]{\includegraphics[width=0.3\textwidth]{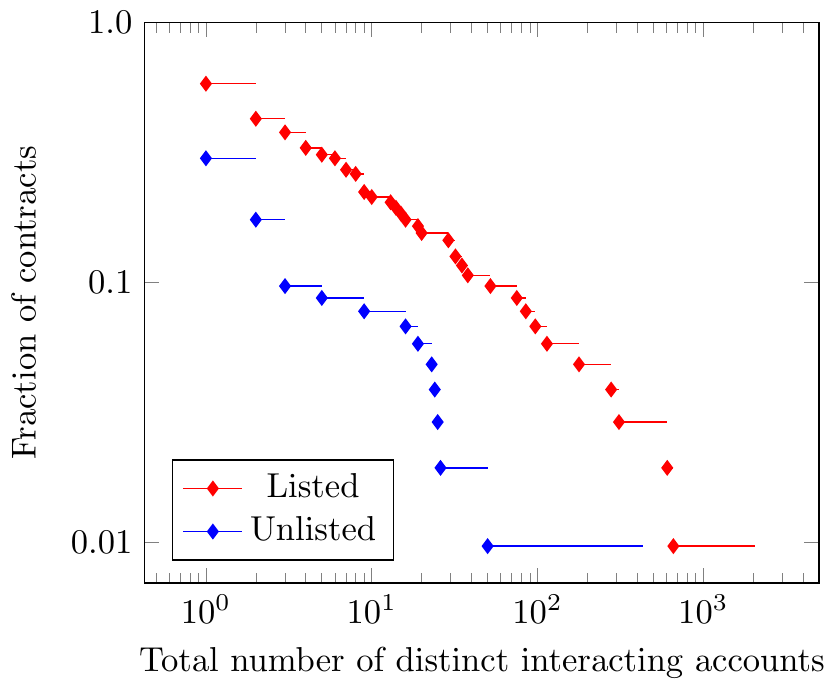}\label{fig:nDistinct_ecdf}}
\caption{Empirical Complementary Cumulative Distribution Function (CCDF) plots of (a) the total number of sent transactions, (b) the total amount of sent ether, and (c) the total of distinct connecting addresses of the unlisted (blue) and listed (red) smart contracts respectively. The CCDF denotes the fraction of contracts that have more transactions than the corresponding $x$-value. High values of the CCDF mean that a contract type is more successful. The $x$-axis and $y$-axis are on a logarithmic scale. For this sample we used the contracts mined in October 2018.}
\label{fig:tx_ccdfs}
\end{figure*}

For the month October 2017, we were able to match 181 listed smart contracts to an unlisted contract and for October 2018 we found 103 matched pairs. \autoref{fig:tx_ccdfs} shows for both the listed and the unlisted contracts of October 2018 the empirical Complementary Cumulative Distribution Functions (CCDFs) of the three key metrics. In an $x$-$y$ plot, a CCDF denotes on the $y$-axis the fraction of contracts that have more transactions than the corresponding $x$-value. For example, in Figure~\ref{fig:nTx_ecdf} it can be seen that for the unlisted contracts, the CCDF at and before $x=10$ (blue line) is slightly over $0.1$. It fact, it is close to $0.136$, which means that less than $13.6\%$ of the unlisted contracts mined in October 2018 had more than 10 transactions, whereas around $86.4\%$ have fewer. By contrast, for the listed contracts the CCDF at $x=10$ (red line) is close to $0.4$ which means that roughly $40\%$ of the listed contracts had more than $10$ transactions. Close to $6.7\%$ of the listed contracts even have more than $1000$ transactions, whereas the faction of unlisted contracts that had more than $1000$ transactions is zero (the highest value is $983$). Hence, \autoref{fig:tx_ccdfs} suggests that the listed contracts are considerably more successful than the unlisted contracts in terms of the number of sent transactions. This difference is also evident for the number of distinct addresses, but less so for the amount of Ether. The CCDF plots are displayed on logarithmic axes --- this facilitates investigation of the tail behaviour of the distributions of the metrics. We can observe in Figures~\ref{fig:nTx_ecdf}~and~\ref{fig:nDistinct_ecdf} that in a log-log plot, the graphs initially behave as a straight line, which is consistent with power-law distributions, before tilting downwards, which is more consistent with log-normal distributions. This is consistent with the findings of  \cite{Pinna2019} (Figures 5 and 6) and \cite{Tonelli2018} (Figures 5-10).

\paragraph*{The differences are statistically significant} The differences between listed and unlisted contracts can also be made more explicit through statistical tests that are in implemented in the statistical tool R.\footnote{\url{https://www.r-project.org/}} In our case, the most appropriate statistical test is the paired Wilcoxon signed-rank test (from the ``stats'' package). Its null hypothesis states that the median difference between a listed contract's metric and its paired equivalent is less than zero, which is precisely what we seek to reject. The test is non-parametric --- i.e., it makes no assumptions about the probability distribution of the sample (e.g., normality in the case of a two-sample $t$-test). The test statistic is computed using the rankings of the pairs in the sample.

\autoref{tab:test_results} shows the test statistics and $p$-values as computed by  R. Although the computed test statistics are approximate, the $p$-values (``probabilities'' that the data is observed given the null hypothesis) are negligible and we can conclude that there is a significant difference between listed and unlisted contracts in terms of the number of transactions and the number of distinct addresses. However, for the amount of ether the Wilcoxon signed-rank test is able to reject the null hypothesis only at a significance level of slightly below $2.5\%$, so here the difference is less pronounced.

\begin{table*}[!ht]
	\centerfloat
	\begin{tabular}{L{0.1\textwidth}L{0.12\textwidth}L{0.1\textwidth}L{0.12\textwidth}L{0.1\textwidth}L{0.12\textwidth}} \toprule
 \multicolumn{2}{c}{\# transactions} & \multicolumn{2}{c}{sent ether} & \multicolumn{2}{c}{\# addresses}\\
Statistic	& $p$-value & Statistic	& $p$-value & Statistic	& $p$-value \\ \midrule
 $2957.5$ & $1.4 \cdot 10^{-6}$ & $271.5$ & $0.02441$ & 1718 & $1.2\cdot10^{-5}$ \\ \bottomrule
	\end{tabular}
	\caption{Summary of the Wilcoxon signed-rank test for October 2018 ($N=103$).}
	\label{tab:test_results}
\end{table*}

To investigate whether these results are not specific to a single month, we conducted the above test for the three metrics in each of the eight months mentioned previously. The results are displayed in \autoref{tab:test_results_all}. We observe that for the number of transactions and the number of distinct addresses, we are able to reject at the 5\% level (and nearly always even at the $0.05\%$ level) when the sample size is sufficiently large (i.e., above 80). For the amount of ether, the difference is less pronounced in all months, although we are still able to reject at the $5\%$ level in all cases where the sample size is sufficient. Hence, there is sufficient evidence to accept hypotheses 1a, 1b, and 1c of \autoref{sec:research_questions}, although the evidence for 1b is not as strong as for the others.

\begin{table*}[!ht]
	\centerfloat
	\begin{tabular}{lc|C{0.2\textwidth}C{0.2\textwidth}C{0.2\textwidth}} \toprule
	Month		& $N$	& \# transactions		& sent ether			& \# addresses \\ \midrule
	Jan 2017	& $13$	& \cellcolor{orange!30}$0.009$	& $0.06$			& \cellcolor{orange!30}$0.003$\\
	Apr 2017	& $16$	& $0.192$			& $0.81$			& $0.244$\\
	Jul 2017	& $83$	& \cellcolor{red!30}$\approx0$	& \cellcolor{orange!30}$0.004$	& \cellcolor{red!30}$\approx0$\\
	Oct 2017	& $181$	& \cellcolor{red!30}$\approx0$	& \cellcolor{orange!30}$0.001$	& \cellcolor{red!30}$\approx0$\\
	Jan 2018	& $109$	& \cellcolor{red!30}$\approx0$	& \cellcolor{red!30}$\approx0$	& \cellcolor{red!30}$\approx0$\\
	Apr 2018	& $152$	& \cellcolor{red!30}$\approx0$	& \cellcolor{orange!30}$0.003$	& \cellcolor{red!30}$\approx0$\\
	Jul 2018	& $155$	& \cellcolor{red!30}$\approx0$	& \cellcolor{red!30}$\approx0$	& \cellcolor{red!30}$\approx0$\\
	Oct 2018	& $103$	& \cellcolor{red!30}$\approx0$	& $0.37$			& \cellcolor{red!30}$\approx0$\\
	\bottomrule
	\end{tabular}
	\caption{Summary of the Wilcoxon singed-rank test's $p$-values for eight different months. The entries in orange were significant at the $5\%$ level, whereas the entries in red were significant even at the $0.05\%$ level.}
	\label{tab:test_results_all}
\end{table*}

\begin{table*}[!ht]
	\centerfloat
	\begin{tabular}{L{0.2\textwidth}L{0.4\textwidth}R{0.1\textwidth}R{0.1\textwidth}} \toprule
	\textbf{Independent variables}	& \textbf{Choices}	& \textbf{Frequency}	& \textbf{Percentage} \\ \midrule
	Age ($N=218$)			& $< 20$ years		&   4			&  1.8\% \\
					& $\geq 20$ and $< 50$ years
								& 208			& 95.4\% \\
					& $>50$ years		&   6			&  2.8\% \\ \midrule
	Sex ($N=215$)			& Female		&  11			&  5.1\% \\
					& Male			& 204			& 94.9\% \\ \midrule
	Continent ($N=195$)		& Oceania		&   3			&  1.5\% \\
					& East Asia (without mainland China)
								&  16			&  8.2\%\\
					& South Asia (mostly India)
								&  53			& 27.2\% \\
					& Russia, Belarus \& Ukraine
								&  14			&  7.2\% \\
					& Middle East		&   7			&  3.6\% \\
					& Europe		&  65			& 33.3\% \\
					& Africa		&   5			&  2.6\% \\
					& North America		&  22			& 11.3\% \\
					& Middle \& South America
								&  10			&  5.1\% \\ \bottomrule
	Education ($N=217$)		& No degree		&  27			& 12.4\% \\
					& UG degree		&  96			& 44.2\% \\
					& PG degree		&  94			& 43.3\% \\ \midrule
	Main role ($N=221$)		& Engineering		& 116			& 52.5\% \\
					& Business		&  57			& 25.8\% \\
					& Education/Research	&  48			& 21.7\% \\ \midrule
	Years of software		& $< 2$ years		&  44			& 19.9\% \\
	development experience		& $\geq 2$ and $< 5$ years
								&  67			& 30.3\% \\
	($N=221$)			& $\geq 5$ years	& 110			& 49.8\% \\ \midrule
	Job type ($N=219$)		& Self-employed		& 101			& 46.1\% \\
					& Company		& 103			& 47.0\% \\
					& Other			&  15			&  6.8\% \\ \midrule
	Company size ($N=203$)		& 1-4 employees		&  66			& 32.5\% \\
					& 5-49 employees	&  78			& 38.4\% \\
					& $\geq$ 50 employees	&  59			& 29.1\% \\ \bottomrule
	\end{tabular}
	\caption{Descriptive statistics  of the demographics}
	\label{tab:demographics}
\end{table*}

\subsection{Questionnaire analysis}\label{sec:qnr-analysis}
In the months April and May of 2019 we received 376 responses. We discarded 147 incomplete responses, and 4 responses that had selected the same option for most questions. This left us with 225 usable responses. Most questions had a relatively large range of choices. For example we provided 7 choices for the highest level of education attained but it turned out that 3 choices would have been enough. We recoded the data to combine choices that were rarely selected with more common choices.

\paragraph*{Independent variables} The independent variables and the recoded choices, as well as the descriptive statistics are summarised in \autoref{tab:demographics} for the demographics and in \autoref{tab:ethereum} for the Ethereum-related questions.

\paragraph*{Age and Sex} The respondents are relatively young and predominantly male. This is probably a reflection of the general situation in the IT industry.

\paragraph*{Continent} Respondents hail from all continents. The country with the largest representation is India. This is probably due to the fact that since the 1980s many IT jobs, particularly from the US, have been outsourced to India. It has therefore gained a strong market position in IT.
Ethereum appears to be more prominent as a technology in Europe and India than in the USA.
None of the respondents have indicated that they are from mainland China.
The reasons for this are unclear: the Chinese user base is not overwhelmingly smaller than its Indian counterpart --- 44 million versus 53 million, respectively\footnote{\url{https://www.businessofapps.com/data/linkedin-statistics/}} --- and a widening of the search queries to include Chinese script (e.g., \raisebox{-0.1cm}{\includegraphics[height=\baselineskip]{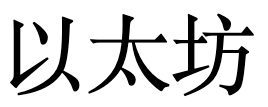}} for Ethereum) did not lead to signifcantly more results.

\paragraph*{Education and experience}%
\phantomsection%
\label{education_and_experience}%
The respondents are well educated and they are experienced in software development, but less experienced in development on Ethereum. This is to be expected, as the technology is relatively young. Ethereum offers many novel features, for example that every computation costs gas (i.e., real money). Learning how to use Ethereum properly is therefore probably more difficult than learning to use a mainstream language, such as Java or Python. A steep learning curve combined with time to market pressure provides a significant challenge to the developer.

\paragraph*{Main role}%
\phantomsection%
\label{main_role}%
The majority (52.5\%) of the respondents are developers, either of the Ethereum technology itself or of the smart contracts and DApps using Ethereum. The second largest group of respondents (25.8\%) has business roles, such as investor, CEO, CTO, business developer, marketer, and project manager. More than one third (36.6\%) of the respondents combine two or more roles. The vast majority of the respondents were recruited via LinkedIn, and all these respondents list Ethereum and Solidity as their skills. Therefore even the respondents in business roles have technical skills.

\paragraph*{Job type and company size}%
\phantomsection%
\label{small_teams}%
Most respondents work in small companies, probably start-ups or scale-ups, which matches the general perception of the blockchain industry as a young industry. And because of this, the majority of the respondents work on other tasks beyond Ethereum. Almost half the respondents are self-employed. Some respondents indicated that they are part-time self-employed, and that they have a regular job unrelated to Ethereum.

\begin{table*}[!ht]
	\centerfloat
	\begin{tabular}{L{0.2\textwidth}L{0.4\textwidth}R{0.1\textwidth}R{0.1\textwidth}} \toprule
	\textbf{Independent variables}	& \textbf{Choices}	& \textbf{Frequency}	& \textbf{Percentage} \\ \midrule
	Years of development		& $< 1$ year		&  62			& 28.2\% \\
	experience on Ethereum		& $\geq 1$ and $< 2$ years
								&  83			& 37.7\% \\
	($N=220$)			& $\geq 2$ years	&  75			& 34.1\% \\ \midrule
	Job percentage devoted		& $< 20\%$		&  89			& 41.2\% \\
	to Ethereum  ($N=216$)		& $\geq 20\%$ and $< 50\%$		
								&  49			& 22.7\% \\
					& $\geq 50\%$		&  78			& 36.1\% \\ \midrule
	Usual site for making		& Etherscan		& 161			& 71.6\% \\
	listings ($N=225$)		& Other (e.g. StateoftheDapps, github)	
								&  64			& 28.4\% \\ \midrule
	Total number of			& No listings		&  66			& 29.5\% \\
	listings made ($N=224$)		& 1-4 listings		&  92			& 41.1\% \\
					& $\geq 5$ listings	&  66			& 29.5\% \\ \midrule
	Months between making		& $< 1$ month		&  30			& 40.5\% \\
	the last listing and today	& $\geq 1$ and $< 2$ months		
								&  12			& 16.2\% \\
	($N=74$)			& $\geq 2$ months	&  32			& 43.2\% \\ \midrule
	Application category 		& Development (of Ethereum and its tools)
								&  26			& 13.6\% \\
	that describes best the 	& Finance (e.g., wallets, exchanges, ICO)
								&  46			& 24.1\% \\
	last listed contract 		& Utility (e.g., education, health, energy, identity, supply chain)
								&  59			& 30.9\% \\
	($N=191$)			& Pleasure (e.g., gaming, gambling)
								&  42			& 22.0\% \\
					& Other (e.g., advertising, conservation)
								&  18			&  9.4\% \\ \midrule
	Number of listings made		& $1$ listing		&  51			& 26.8\% \\
	on usual site ($N=190$)		& 2-4 listings		&  71			& 37.4\% \\
					& $\geq 5$ listings	&  68			& 35.8\% \\ \midrule
	Tools used, e.g., mythX,	& No tools		&  38			& 25.7\% \\
	securify ($N=148$)		& $1$ tool		&  62			& 41.9\% \\
					& $\geq 2$ tools	&  48			& 32.4\% \\ \bottomrule
	\end{tabular}
	\caption{Descriptive statistics for the Ethereum-related questions}
	\label{tab:ethereum}
\end{table*}

\begin{table}[!ht]
	\centerfloat
	\begin{tabular}{L{0.2\textwidth}|L{0.2\textwidth}} \toprule
\textbf{Dependent Variables}		& \textbf{Success indicators} \\ \midrule
	Perceived Usefulness		& Incoming Transactions \\
	Ease of use			& Ether received \\
	Listing improves success	& Number of unique interacting addresses \\
	Attention improves success	& Other, please specify \\ \bottomrule
	\end{tabular}
	\caption{Dependent variables and success indicators}
	\label{tab:dependents}
\end{table}

\paragraph*{Listing} The majority of the respondents use Etherscan to list smart contracts. A minority also uses StateoftheDapps, and repositories, such as GitHub. Some respondents work on private block chains (such as Quorum\footnote{\url{https://consensys.net/quorum/}}), and therefore do not list on public websites. The majority has experienced the listing process recently.

\paragraph*{Application category}%
\phantomsection%
\label{application_category}%
Smart contracts and DApps are used for many different applications. The main categories are Finance and Entertainment, which is unsurprising. However, a large segment of the industry works in energy, health, identity management and other utilitarian topics. Probably all application areas are represented. We think that developers may sometimes be trying to develop new services with blockchain technology, even where traditional database technology might be able to achieve the same, or perhaps even better results~\cite{Chowdhury2018,Wuest2018}.

\paragraph*{Development tools}%
\phantomsection%
\label{development_tools}%
The majority of the respondents use, or have used sophisticated experimental tools for Ethereum, such as Echidna\footnote{\url{https://github.com/trailofbits/echidna}}, Manticore\footnote{\url{https://github.com/trailofbits/manticore}}, MythX\footnote{\url{https://mythx.io}} (formerly Mythril), Oyente\footnote{\url{https://github.com/melonproject/oyente}}, Securify\footnote{\url{https://securify.chainsecurity.com}}, SmartCheck\footnote{\url{https://tool.smartdec.net}}, Slither\footnote{\url{https://github.com/crytic/slither}}, and Maian\footnote{\url{https://github.com/MAIAN-tool/MAIAN}}.
However, many respondents indicated that the current development tools, such as the Truffle framework Framework\footnote{\url{https://www.truffleframework.com}}  and Remix\footnote{\url{https://remix.ethereum.org}} are more important than the sophisticated experimental tools, and that the standard tools have enough shortcomings already. We take this as an indication that there is a lack of maturity in the tools that are available.

\paragraph*{Dependent variables}
There are five dependent variables as summarised in \autoref{tab:dependents}. ``Perceived usefulness'', and ``Ease of use'' are classical TAM constructs based on 6 items each, as described in \autoref{sec:method}. ``Listing improves success'', and ``Attention improves success'' are new constructs inspired by NPD, which are intended to measure to what extent the respondent thinks that using a particular technology improves the success of a smart contract. ``Listing improves success'' is based on 9 items that all explicitly mention in the question what the respondent indicated as his/her favourite listing site, which, in most cases, is Etherscan. ``Attention improves success'' is intended to capture four important activities that can all contribute to the success of smart contracts.
The last dependent variable captures what the respondent believes is the best measure for success of a smart contract. We gave a number of choices as indicated in \autoref{tab:dependents}, but respondents were also free to suggest their own.

\begin{table*}[!ht]
	\centerfloat
	\begin{tabular}{L{0.17\textwidth}R{0.1\textwidth}R{0.1\textwidth}R{0.1\textwidth}R{0.1\textwidth}R{0.1\textwidth}R{0.1\textwidth}} \toprule
						& Strongly disagree	
								& Disagree	& Neutral	& Agree		& Strongly agree& Cronbach's $\alpha$ \\ \midrule
	Perceived usefulness (N=199)		& \bf 1.5\%	& \bf 7.0\%	& 31.2\%	& 49.7\%	& 10.6\%	& 0.94 \\[0.1cm]
	Ease of use (N=201)			&  0.5\%	&  4.5\%	& 31.8\%	& 51.7\%	& 11.4\%	& 0.91 \\[0.1cm]
	Listing improves success (N=193)	&		&  5.2\%	& \bf 54.4\%	& 38.3\%	&  2.1\%	& 0.87 \\[0.1cm]
	Attention improves success (N=213)	&		&  1.9\%	&  8.9\%	& \bf 60.1\%	& \bf 29.1\%	& 0.73 \\ \bottomrule
	\end{tabular}
	\caption{Respondents' scores on the four scales (highest percentage per column in bold face).}
	\label{tab:four_scales}
\end{table*}

\paragraph*{Reliability} The statistics of the first four dependent variables are shown in \autoref{tab:four_scales}. In each case, Cronbach's $\alpha$ is high, and the correlation between the items are all positive and high, so that we can conclude that all scales are reliable.

\paragraph*{Perceived usefulness and ease of use} The majority of the respondents consider listing services useful. This is to be expected as using a listing service boils down to completing a form on the web. The null hypothesis that a majority of developers do not agree that listing is useful can be rejected at the $5\%$ level, as the critical value for a binomial distribution with $N=199$ and $p=\frac{1}{2}$ is 88 and only 79 respondents ($39.7\%$) did not agree (in fact, the $p$-value is only $0.22\%$). Since most respondents have actually used listing services, it is not a surprise that they find such services useful. The null hypothesis that a majority of developers do not agree that listing services are easy to use, with 74 out of 201 not agreeing ($p\approx0.011\%$). Hence, we have enough evidence to support hypotheses 2a and 2b of \autoref{sec:research_questions}.

\begin{table*}[!ht]
	\centerfloat
	\begin{tabular}{p{0.3\textwidth}R{0.1\textwidth}R{0.1\textwidth}R{0.1\textwidth}R{0.1\textwidth}R{0.1\textwidth}} \toprule
	\multicolumn{6}{c}{Listing improves success ($\chi^2=15.5$, $p=0.016$)} \\
	Highest education level		& Strongly disagree
							& Disagree	& Neutral	& Agree		& Strongly agree \\ \midrule
	No degree (N=23)		&		& \bf 8.7\%	& 43.5\%	& 39.1\%	& \bf 8.7\% \\[0.1cm]
	UG degree (N=85)		&		& 1.2\%		& 49.4\%	& \bf 48.2\%	& 1.2\% \\[0.1cm]
	PG degree (N=81)		&		& 6.2\%		& \bf 64.2\%	& 28.4\%	& 1.2\% \\[0.1cm]
	Total (N=189)			&		& 4.2\%		& 55.0\%	& 38.6\%	& 2.1\% \\ \bottomrule \toprule
	\multicolumn{6}{c}{Attention improves success ($\chi^2=12.9$, $p=0.044$)} \\
	Highest education level		& Strongly disagree
							& Disagree	& Neutral	& Agree		& Strongly agree \\ \midrule
	No degree (N=25)		&		& \bf 4.0\%	& \bf 12.0\%	& 48.0\%	& 36.0\% \\[0.1cm]
	UG degree (N=91)		&		& 2.2\%		& 7.7\%		& 52.7\%	& \bf 37.4\% \\[0.1cm]
	PG degree (N=90)		&		&		& 7.8\%		& \bf 73.3\%	& 18.9\% \\[0.1cm]
	Total (N=206)			&		& 1.5\%		& 8.3\%		& 61.2\%	& 29.1\% \\ \bottomrule
	\end{tabular}
	\caption{The highest level of education versus the claims that ``listing improves success'' and ``attention improves success'' (highest percentage per column in bold face).}
	\label{tab:education}
\end{table*}

\paragraph*{Listing improves success and attention improves success}%
\phantomsection%
\label{listing_is_not_enough}%
\autoref{tab:education} shows that the majority of the respondents neither agree nor disagree with the proposition that ``listing improves success''.
In fact, only 78 out of 193 respondents agree, and the null hypothesis that they form a majority can be rejected at the $5\%$ level ($p\approx0.47\%$). Several respondents noted in the open questions that listing is usually done as part of a broader package of measures, and that listing on its own does not improve success. Hence, hypothesis 2c of \autoref{sec:research_questions} is not supported by evidence.

Before a marketing campaign begins, the developers will have made sure that the contract and/or the DApp has been thoroughly tested and passed a security audit. The contracts may even have been formally verified and listed on GitHub. All these activities foster trust in the contract, and we have captured all these under ``Attention improves success''. Indeed the vast majority (190 out of 213) of the respondents agree that devoting attention to the contract improves its success, and the null hypothesis that they do not agree can be rejected with overwhelming confidence ($p\approx3.5\cdot10^{-34}$).

\autoref{tab:education} also shows that respondents with a higher education degree are more sceptical than other respondents about the ability of the listing process to improve success. All respondents are more positive about devoting attention to a contract that to just listing it. The difference in both cases is statistically significant.

We found no significant differences between respondents in engineering or business roles. We take this as an indication that engineering and business have a shared view on how to make a smart contract successful.

\begin{table*}[!ht]
	\centerfloat
	\begin{tabular}{L{0.3\textwidth}R{0.1\textwidth}R{0.1\textwidth}R{0.1\textwidth}R{0.1\textwidth}R{0.1\textwidth}} \toprule
	\multicolumn{6}{c}{More successful if they have been open sourced ($\chi^2=15.0$, $p=0.059$)} \\
	Years of Ethereum experience	& Strongly disagree
							& Disagree	& Neutral	& Agree		& Strongly agree \\ \midrule
	1 year (N=59)			& \bf 1.7\%	& 3.4\%		& \bf 25.4\%	& \bf 50.8\%	& 18.6\% \\[0.1cm]
	$\geq 1$ and $< 2$ years (N=77)	&		&		& 15.6\%	& 41.6\%	& \bf 42.9\% \\[0.1cm]
	$\geq$ 2 years (N=74)		&		& \bf 4.1\%	& 20.3\%	& 36.5\%	& 39.2\% \\[0.1cm]
	Total (N=210)			& 0.5\%		& 2.4\%		& 20.0\%	& 42.4\%	& 34.8\% \\ \bottomrule \toprule
	\multicolumn{6}{c}{More successful if they have been formally verified ($\chi^2=17.1$, $p=0.029$)} \\
	Years of Ethereum experience	& Strongly disagree
							& Disagree	& Neutral	& Agree		& Strongly agree \\ \midrule
	1 year (N=59)			& \bf 1.7\%	& 1.7\%		& 10.2\%	& \bf 72.9\%	& 13.6\% \\[0.1cm]
	$\geq 1$ and $< 2$ years (N=81)	& 1.3\%		& 1.3\%		& 12.7\%	& 57.0\%	& \bf 27.8\% \\[0.1cm]
	$\geq$ 2 years (N=77)		& 		& \bf 8.0\%	& \bf 22.7\%	& 50.7\%	& 18.7\% \\[0.1cm]
	Total (N=217)			& 0.9\%		& 3.8\%		& 15.5\%	& 59.2\%	& 20.7\% \\ \bottomrule
	\end{tabular}
	\caption{The number of years of development experience on Ethereum versus the claims that ``open sourcing improves success'' and ``formal verification improves success'' (highest percentage per column in bold face).}
	\label{tab:experience}
\end{table*}

\autoref{tab:experience} shows that respondents who have more experience with development on Ethereum are more sceptical about measures to promote success. The difference is significant in both cases. No significant differences were found for the two other claims (i.e. smart contracts are more successful if they have passed a security audit, and smart contracts are more successful if they have been listed). This does not mean that the respondents consider listing unimportant. Instead it means that respondents see listing as part of a package and as we saw before, respondents are mostly neutral about the effect of listing alone (see, e.g., \autoref{tab:education}).

We found a small but significant difference between respondents in engineering and business roles. The engineers are slightly more sceptical about the benefits of formal verification, whereas the respondents in a business role are more sceptical about open sourcing. Apparently, the familiarity with an activity and its effect correlates with scepticism about the benefits.

\subsection{Open questions}
In the last question of the survey we asked respondents for anything else that they wish to share with us. Many did so and we summarise their remarks here. We group the responses by topic.

\paragraph*{Management}
Some respondents note that an on-chain registry of projects, and a rating system like that for mobile apps would be useful. At present users have no reliable information on new projects, which does not foster trust in the eco-system.
An identity management system could help end-users to differentiate between bona-fide DApps and scams. This is all the more important, as there is almost no legal support for scam victims.
An insurance system would be welcome, to compensate scam victims.

\paragraph*{Language}
\phantomsection%
\label{language}%
Solidity, the most commonly used programming language contains many innovations, but also obscure features that respondents are struggling with. As one respondent put it:
\begin{quote}
An alternative for solidity that has less gotcha's for smart contract developers. Preferable together with an IDE that has good formal verification support that is usable for non-academic users and has a plugin system that allows for a community to develop and share plugins that can verify or optimize a smart contract code.
\end{quote}
Better support for Vyper, which is an alternative for Solidity, would also be welcome.

\paragraph*{Interfacing}
Estimating the amount of gas needed for transactions is too much of an art.
There is a need for integrated support of distributed storage, as it is clearly impractical to store more than essential data on the blockchain itself.
An intuitive API generator for smart contracts is needed to facilitate developing DApps.
The mobile experience of DApps should be improved.

\paragraph*{Tooling}%
\phantomsection%
\label{tooling}%
A better user interface is needed to visualise smart contract statistics.
Better blockchain explorers are also needed, with more and better filters to be able to focus on the data that matters.
Many respondents remark that current integrated development environments (IDE) are lacking a symbolic debugger or the ability to check the state of a contract at a certain block etc. IDEs also lack support for linting, testing, fuzzing, contract updates, gas estimation, DApp interfacing, and performance monitoring.
Several respondents note that testing tools, such as fuzzers and mutators are desperately needed.

\paragraph*{Training}
It is not easy to learn how to work with Ethereum, or its current tools. The learning curve is steep, and the documentation is scattered.
Most of the tooling is not particularly user friendly.
Several respondents note that audits and code reviews are a challenge because auditors are insufficiently familiar with Ethereum.

%LOREM IPSUM LOREM IPSUM LOREM IPSUM LOREM IPSUM
%LOREM IPSUM LOREM IPSUM LOREM IPSUM LOREM IPSUM
%LOREM IPSUM LOREM IPSUM LOREM IPSUM LOREM IPSUM
%LOREM IPSUM LOREM IPSUM LOREM IPSUM LOREM IPSUM

\section{Recommendations}
\label{sec:recommendations}
We provide 10 recommendations for: 
smart contract application developers (denoted by $\mathbb{A}$), 
core developers of smart contract platforms (denoted by $\mathbb{C}$), and  developers/designers/maintainers of auxiliary infrastructure (such as listing services) for smart contract platforms, which we refer to as moderators (denoted as $\mathbb{M}$).
We have validated these recommendations via interviews with an international panel of senior Ethereum smart contract developers as well as a few junior engineers, and incorporated their feedback. 
We further refer to these validating developers as validators.
The recommendations are split into three categories according to their scope, while in each category we explicitly state the target audience by one of letters  $\mathbb{A}$, $\mathbb{C}$, and $\mathbb{M}$.

%\ih{We implicitly assume the target audience are smart contract developers. When the target audiences are core developers or moderators, we explicitly state it.}.

\subsection{Eco-system}
The first set of recommendations addresses the eco-system of companies, universities, and governments, who all share a responsibility for success of innovative technologies based on blockchains.

\textbf{Improve Ethereum experience $\{\mathbb{A}\}$} Most Ethereum developers are well-educated, but have limited development experience on Ethereum (See \autoref{education_and_experience}, page~\pageref{education_and_experience}).
Therefore, developers may be unaware of best practices for smart contract programming\footnote{\url{https://consensys.github.io/smart-contract-best-practices/}} or security vulnerabilities and threats specific to smart contracts.\footnote{\url{https://github.com/SmartContractSecurity/SWC-registry}}
As confirmed by our validators, the most important recommendation for beginning Ethereum developers is to team up with more experienced developers to transfer knowledge and skills.
Further, universities may wish to offer a blockchain engineering courses to IT engineering students, who will be the next generation of developers.

% \phh{Nick did not like the hackathons?}
% Another option to gain experience at an accelerated pace are hackathons, capture-the-flag challenges, and other smart contract programming competitions -- they are usually organized by experienced developers.

\textbf{Promote native integration of development tools $\{\mathbb{M}\}$ } 
The most popular integrated development environments (IDEs), such as Visual Studio Code and IDE,\footnote{\url{https://code.visualstudio.com/} and \url{https://visualstudio.microsoft.com/}} NetBeans,\footnote{\url{https://netbeans.org/}} and Eclipse,\footnote{\url{https://www.eclipse.org/ide/}} do not natively support Ethereum development tools (See \autoref{tooling}, page~\pageref{tooling}).
In most of these IDEs, it is possible to install a third-party plugins to support Ethereum development.
However, such plugins might be buggy and do not provide sufficient programming experience.
Hence, we encourage the maintainers of popular IDEs to consider native integration for Ethereum smart contract development, especially when taking into account the rise of potential for blockchain-based applications as well as their proliferation.

%In the traditional IT industry, large companies with extensive experience, such as Google, Apple, Microsoft, and IBM generally make more reliable products than small companies with limited experiences and small teams (See \autoref{tooling}, page~\pageref{tooling}). 
%It is advisable to look for a combination of the strengths of young and dynamic companies with the experience of large corporations.
 
% \phh{Netbeans footnote too much of a detail.}
% \footnote{For example in the case of NetBeans, there is not even a third party plugin available.}

\textbf{Check before listing $\{\mathbb{A, M}\}$} 
As indicated by the respondents of our questionnaire, listing is usually part of a marketing campaign.
However, our respondents also indicated that this is only a small part of the whole effort made to achieve business success (See \autoref{listing_is_not_enough}, page~\pageref{listing_is_not_enough}).
The positive impact of the listing on the success of the business was also confirmed from our statistical analysis of the questionnaire's data, where we found a correlation of these two phenomena.
Nevertheless, we extend the recommendation for listing by a recommendation to perform extra checks and automatic scrutiny.
For example, listing sites could run static analysers (See \autoref{development_tools}, page~\pageref{development_tools}) as part of the listing process, and propose improvements to the code before deployment.

\subsection{Business}
The second set of recommendations focuses on the business perspective.

\textbf{Put the business in the lead $\{\mathbb{A}\}$}  
In general, to achieve product advantage, a product must~\cite{Cooper1987}: 
(1) be better than comparable products, 
(2) solve a problem for users, 
(3) be innovative, 
(4) reduce costs, and 
(5) offer unique benefits.
Most of these items require a deep understanding of the underlying business as well as the technology (See \autoref{main_role}, page~\pageref{main_role}).  It is therefore advisable to give the business a leading role, and let the technology follow the business.
For example, the Video 2000 standard for videotapes was technically better than competing standards, but it entered the market too late to become a success~\cite{Johne1994}.

\textbf{Foster in-company collaboration $\{\mathbb{A}\}$} 
Developers often work alone, or part-time, or perform more roles at once, and lack the skills and resources of a larger coherent team that includes members with marketing experience as well (See \autoref{small_teams}, page~\pageref{small_teams}).
As confirmed by our validators, it is important that all members of the company (including marketing specialists) understand the theoretical background of the blockchain technology and the reason for its usage.
Hence, it is advisable to work in more specialized teams that focus on particular domain but closely collaborate together.
A few validators confirmed that outsourcing of smart contract development is also on the rise, while the main company focuses more on the business and development of the user interface -- this indirectly witnesses that specialized teams might bring higher productivity and success.
Further, we encourage to use community means\footnote{\url{https://ethereum.stackexchange.com/}} for discussion and resolution of both technical and business issues.

% \phh{Reformulated to be consistent with the other recommendations as something the can be acted upon}
\textbf{Improve information on the quality of products $\{\mathbb{M, A, C}\}$} 
More than 10,000 smart contracts are added to the Ethereum blockchain every day.
Most of them will never even be used, and some of them might be scams (See \autoref{detailed_comparison}, page~\pageref{detailed_comparison}).
It is advisable to investigate how potential users can be better informed about the quality and popularity of the offerings.
For example, DApp directories such as StateoftheDapps already provide a ranking system. 
However, any ranking system is as good as the data on which it is based, therefore data curating (i.e., processing and filtering of misinformation) is vital.
Since a ranking is often realized by centralized parties (e.g. StateoftheDapps, ForkDelta), it might be potentially biased. 
Hence, also means of decentralization might be considered for fair ranking either as a decentralized application or a natively supported service of the smart contract platform.
Another consideration for ranking is to distinguish rankings made by developers with known identities and rankings made by the users of smart contract applications -- each of them might indicate different features (i.e. popularity, soundness, optimality).

\textbf{Use blockchain technology only when you really need it $\{\mathbb{A}\}$} Blockchains have unique properties, such as distributed control, transparency, immutability, availability, and censorship resistance.
It is crucially important to build on top of such features in order to make a business successful. (See \autoref{application_category}, page~\pageref{application_category})
For example, if an application can be built using standard database technology, building it on a blockchain is unlikely to lead to product advantage because the DApp will be more expensive to use and have worse performance.

\subsection{Engineering}
The last set of recommendations is specific to engineers.
	
\textbf{Improve languages and toolsets $\{\mathbb{C, M}\}$} Mainstream programming languages such as Python and Java have evolved over many years.
Programming languages for smart contracts are still in their infancy (See \autoref{language}, page~\pageref{language}).
We hope that experienced language designers are willing to develop a better next generation of smart contract languages. At the same time, efforts to integrate WebAssembly standards into Ethereum\footnote{\url{https://github.com/ewasm/design}} (ewasm) are promising.
For increased safety, there is a trend to use Turing-incomplete smart contract languages (e.g., Scilla and Pact), although this provides limited expressiveness and hence limited applicability.
Another direction in this category is to develop source code optimizers that should explicitly focus on savings of the gas consumption. 
This might be further combined with regression testing to support even more experimental and powerful optimizations.

\textbf{Improve testing $\{\mathbb{A, M}\}$} A service can be superior to all its competitors, but if it is unreliable, it will probably not be successful~\cite{Bianco2011} (See \autoref{listing_is_not_enough}, page~\pageref{listing_is_not_enough}).
It is advisable to test a service thoroughly before placing it on the market, which is too difficult with the current tooling. 
A potential direction here is to make automatic unit test generators.

% \phh{the template story requires much more explanation than given here to be understandable}
% (or template test generators\footnote{Note that templates need to be revised and amended by developers.}) that can cover some common and corner cases.

\textbf{Support updates $\{\mathbb{C}\}$} Once it has been deployed on the blockchain, a smart contract can no longer be updated (See \autoref{tooling}, page~\pageref{tooling}).
Current methods of dealing with this are patchwork and it is advisable to look for new solutions.
For example, a developer must make sure that he fully understands the risks of splitting a contract in an immutable ``front-end'' contract and a ``back-end'' library~\cite{Mavridou2019} that can be potentially updated.
To alleviate negative consequences of buggy ``back-end'' updates might be to deploy unit tests along with a smart contract application, while the smart contract platform itself could enforce the new updates of the libraries to pass all unit tests of the ``front-end'' smart contract.

\subsection{Validation}
We validated the proposed recommendations with a few senior developers with several years of experiences (in some cases 10 years) with Ethereum, but also with a few junior developers and students.
In detail, 
All of the application developers $\mathbb{A}$ are running a successful business with Ethereum smart contracts used for multiple products.
Our sample of validators contain four application developers $\mathbb{A}$, two core developers of smart contract platforms $\mathbb{C}$, and one developer focusing on an auxiliary infrastructure of Ethereum smart contract platform $\mathbb{M}$.
From demographic perspective, we remark that five of the validators reside in Asia, one in Europe, and one in Australia. 

In the validation process, first we shared the list of recommendations in a written form with the validators, and then we arranged a phone call for a further discussion.
The validators were tasked to accept or reject the recommendations and also to indicate other directions and issues that they would appreciate to be addressed in this field.
The most significant change that we made in response to the validation was to indicate the appropriate audience for each recommendation: application developers, core platform developers, and/or auxiliary infrastructure developers.
The next significant change was inclusion of recommendation that suggests to promote native integration of development tools in the most popular IDEs.

% Nicolas Addison [A] - Australia, Sydney, Solution Architect of Consensys' project, former CTO
% Kenneth Hu [A] - Asia, Singapore, CTO at BayPay
% Ming Chan [C] - Europe, Switzerland, former Executive Director of Ethereum Foundation
% Kaven Choi [M] - Asia, Malaysia, Ethereum Community Support at Etherscan.io
% Chih-Cheng Liang [C] - Asia, Taiwan, Ethereum Foundation
% <Korean name> [A] - Asia, Korea, developer at Bithump
% William [A] - Asia, Singapore, undergrad student at NTU

%
% \phh{Next remark already mentioned}
% Another interesting trend that the discussion revealed is an outsourcing of the smart contract development by blockchain-oriented companies.

\section{Limitations}
\label{sec:limitations}
In this section we discuss potential threats to the validity of our work.

%\daniel{This website has lot of information about limitations sections in dissertations: \url{http://dissertation.laerd.com/research-quality.php}} \pieter{good idea to use this}

\subsection{Threats to internal validity}
\paragraph{Blockchain analysis} We have performed tests on three metrics for business performance to identify whether listed smart contracts are more successful in practice than unlisted smart contracts. Although these three metrics are not independent, none of the tests that we have applied require that they are. The heuristic for pairing listed and unlisted contracts is based on recently published work \cite{He2019}, and we have validated the heuristic (See  \autoref{tab:nld_cross_table}). 
We have also used the methodology of \cite{Angelo2019} to test if unlisted contracts that have a functionally similar listed contract score differently on the three metrics than those that do not -- if they did, then this would undermine the validity of our results as we would no longer be able to assume that our samples were drawn from the same population. In our sample of 812 pairs, 461 unlisted contracts are functionally similar to a listed contract, but they were not found to be significantly more successful than unlisted contracts that are not functionally similar.

\paragraph{Questionnaire analysis} We have taken steps to avoid experimenter effects that could threaten the validity of our results. To avoid social desirability bias, we did not tell the respondents that we found listed smart contracts to be more successful than unlisted smart contracts. Like all surveys based on self-reporting, it is possible that some respondents have not answered the questions truthfully. We have discarded incomplete surveys, assuming that respondents who took the effort to complete the survey would probably also have made an effort to answer the questions truthfully. Also, the anonymity of the survey has hopefully had a mitigating effect on social desirability bias. 

The questionnaire is focused on the listing process. To obtain a broader picture of the marketing strategy, we could have asked more questions, e.g., about the skills, and resources of management, marketing, customer relations, advertising, and sales. However, we were hesitant to ask such commercially sensitive information, as this may have an adverse effect on the response rate.

\subsection{Threats to construct validity}

\paragraph{Blockchain analysis}

Our proxies for the success metrics --- i.e., sales, profits, and market share  --- can either overestimate, or underestimate those variables. However, this is a consequence of the limitations of the publicly available dataset. Further research would be needed to investigate the relationship between, e.g., market share and the number of distinct number of addresses (for example, some of the addresses could be controlled by the same entity).

\paragraph{Questionnaire analysis}

We pre-tested the questionnaire in several rounds by asking a number of developers and researchers to review it. We use existing constructs from established theory where possible, and the questions in the questionnaire were inspired by existing literature, e.g., \cite{Venkatesh2000} for the questions regarding the perceived usefulness and ease of use of listing sites, and the NPD theory %
% \daniel{Pieter, do you have a good reference for this?}
for the questions regarding a comparison between listed and unlisted contracts. The reliability of the two new constructs that we propose has been assessed by the appropriate statistical tests implemented in the SPSS tool.

\subsection{Threats to external validity}

\paragraph{Blockchain analysis} We have taken uniform random samples of smart contracts from 2 months out of the total of 48 months of data available on the Ethereum blockchain. We repeated the experiment with 3 sets of 2 months and found similar results. This indicates that the results are generalizable to the recent past. Although it was too costly to do all 48 months using Google BigQuery, the months were chosen to reflect all relevant seasons in the two most recent years since the start of our study. We do note that the Ethereum blockchain and the surrounding eco-system (in particular the listing services) have evolved rapidly in those 48 months, and continue to do so.
%For each listed smart contract we searched the blockchain for a similar but unlisted smart contract. 

\paragraph{Questionnaire analysis} Most of our respondents are on LinkedIn, and we assume that LinkedIn membership is representative for the world population of Ethereum developers. Mainland China is not represented. Other surveys including the Stack Overflow Annual Developer Survey\footnote{\url{https://insights.stackoverflow.com/survey/2019}} have apparently also had difficulty reaching out to developers in mainland China. 
To reduce non-response bias, we have approached potential respondents via personal messaging on LinkedIn.

\section{Conclusion}
\label{sec:conclusion}
The transparency and public availability of the blockchain give researchers the opportunity to analyse smart contracts and their transactions in detail. % because all the required data is publicly available.
We have used this transparency to investigate what makes smart contracts successful.
Based on the theory of New Product Development (NPD), we have assumed that: 
(1) the number of transactions per unit of time is a representative proxy for sales, 
(2) the amount of ether sent to a contract per unit of time is decisive for profit, and 
(3) the number of different addresses interacted with a contract per unit of time is indicative of the market share.
According to NPD, these three metrics correlate positively with business success.
To investigate whether this is also the case for smart contracts, we have downloaded a random sample of smart contracts listed on Etherscan (N=812), and we have matched each listed smart contract with a functionally similar contract that is not listed.
In over 80\% of the pairs of similar contracts, the listed contract is more successful than the unlisted contract.
This finding is statistically significant and suggests that ``being listed'' positively correlates with the business success.

We then conducted a survey among mainly LinkedIn members who mentioned Ethereum and Solidity on their profile.
We asked the respondents for their opinion about the feasibility of promoting and measuring the success of smart contracts.
The respondents (N=225) are located on all continents and are well educated.
However, they have relatively little experience with Ethereum and Solidity because the technology is relatively young.
Most respondents do other work in addition to their work with Ethereum, and they either work for their own account or for a small company, rather than for a large organisation.
Most respondents have listed one or more contracts on Etherscan, and find the listing process easy to use.
The majority of respondents consider making a listing as part of a broader package of measures that all contribute to the success of smart contracts.
Respondents in an engineering role are more sceptical about what engineers can do to promote success and less sceptical about what their colleagues in business can do to promote success, and vice versa.  

A common complaint of the respondents is that the Ethereum technology is not yet mature, which is evident from the fact that essential tools such as the debugger have limited functionality and the documentation is fragmented. 
This was one of the main conclusions of the work of Bosu et al.~\cite{Bosu2019}, and now, more than a year later, it is still the most heard complaint.
To address this issue, we have made a dozen recommendations, which we then validated by submitting a draft of the recommendations to a panel of international experts.
The first version of the recommendations appeared to contain ambiguities, which we then clarified based on the interviews, leading to the recommendations of \autoref{sec:recommendations}.
We hope that our recommendations are a useful contribution to the field, and developers may follow or consider some of the recommendations.

We can now answer our research questions as follows:
\begin{itemize}
	\item Listed smart contracts are orders of magnitude more successful than unlisted smart contracts.
	\item Open sourcing and formal verification contribute to success. Listing a smart contract on a suitable website only contributes to success when listing is part of a broader package.
\end{itemize}

\section*{Acknowledgment}
This work was supported in part by the National Research Foundation (NRF), Prime Minister's Office, Singapore, under its National Cybersecurity R\&D Programme (Award No. NRF2016NCR-NCR002-028) and administered by the National Cybersecurity R\&D Directorate.

We thank Susanne Barth for her advice on the questionnaire and we thank Monika di Angelo and Gernot Salzer for labelling our sample of 812 contract pairs with functional similarity indicators.

\bibliographystyle{abbrv}
\bibliography{darkweb_refs,additional_refs}

\begin{thebibliography}{10}

\bibitem{Atuahene1996}
K.~Atuahene-Gima.
\newblock Differential potency of factors affecting innovation performance in
  manufacturing and services firms in {Australia}.
\newblock {\em J. of Product Innovation Management}, 13(1):35--52, Jan 1996.

\bibitem{Atzei2017}
N.~Atzei, M.~Bartoletti, and T.~Cimoli.
\newblock A survey of attacks on {Ethereum} smart contracts ({SoK}).
\newblock In M.~Maffei and M.~Ryan, editors, {\em 6th Conf. on Principles of
  Security and Trust (POST)}, volume 10204 of {\em LNCS}, pages 164--186,
  Uppsala, Sweden, Apr 2017. Springer.

\bibitem{Bartoletti2017a}
M.~Bartoletti and L.~Pompianu.
\newblock An empirical analysis of smart contracts: Platforms, applications,
  and design patterns.
\newblock In {\em Int. Conf. on Financial Cryptography and Data Security (FC)},
  volume 10323 of {\em LNCS}, pages 494--509, Sliema, Malta, Apr 2017.
  Springer.

\bibitem{Bellon2007}
S.~Bellon, R.~Koschke, G.~Antoniol, J.~Krinke, and E.~Merlo.
\newblock Comparison and evaluation of clone detection tools.
\newblock {\em IEEE Transactions on software engineering}, 33(9):577--591, Sep
  2007.

\bibitem{Bosu2019}
A.~Bosu, A.~Iqbal, R.~Shahriyar, and P.~Chakraborty.
\newblock Understanding the motivations, challenges and needs of blockchain
  software developers: a survey.
\newblock {\em Empirical Software Engineering}, 24(4):2636–2673, Aug 2019.

\bibitem{buterin2014next}
V.~Buterin.
\newblock A next-generation smart contract and decentralized application
  platform.
\newblock {\em white paper (2014)}, 2014.

\bibitem{Chowdhury2018}
M.~J.~M. Chowdhury, A.~Colman, M.~A. Kabir, J.~Han, and P.~Sarda.
\newblock Blockchain versus database: A critical analysis.
\newblock In {\em 2018 17th IEEE Int. Conf. On Trust, Security And Privacy In
  Computing And Communications/ 12th IEEE International Conference On Big Data
  Science And Engineering (TrustCom/BigDataSE)}, pages 1348--1353, New York,
  Aug 2018. IEEE.

\bibitem{Cooper1987}
R.~G. Cooper and E.~J. Kleinschmidt.
\newblock New products: What separates winners from losers?
\newblock {\em J. of Product Innovation Management}, 4(3):169--184, Sep 1987.

\bibitem{Datta2013}
D.~Datta and S.~Kajanan.
\newblock Do app launch times impact their subsequent commercial success? an
  analytical approach.
\newblock In {\em Int. Conf. on Cloud Computing and Big Data}, pages 205--210,
  Fuzhou, China, Dec 2013. IEEE.

\bibitem{Bianco2011}
V.~del Bianco, L.~Lavazza, S.~Morasca, and D.~Taibi.
\newblock A survey on open source software trustworthiness.
\newblock {\em IEEE Software}, 28(5):67--75, Sep 2011.

\bibitem{Destefanis2018}
G.~Destefanis, M.~Marchesi, M.~Ortu, R.~Tonelli, A.~Bracciali, and R.~Hierons.
\newblock Smart contracts vulnerabilities: a call for blockchain software
  engineering?
\newblock In {\em Int. Workshop on Blockchain Oriented Software Engineering
  (IWBOSE)}, pages 19--25, Campobasso, Italy, Mar 2018. IEEE.

\bibitem{Angelo2019}
M.~di~Angelo and G.~Salzer.
\newblock Mayflies, breeders, and busy bees in {Ethereum}: Smart contracts over
  time.
\newblock In {\em 3rd ACM Workshop on Blockchains, Cryptocurrencies and
  Contracts (BCC)}, pages 1--10, Auckland, New Zealand, Jul 2019. ACM.

\bibitem{He2019}
N.~He, L.~Wu, H.~Wang, Y.~Guo, and X.~Jiang.
\newblock Characterizing code clones in the {Ethereum} smart contract
  ecosystem.
\newblock Technical report, Beijing University of Posts and Telecommunications,
  May 2019.

\bibitem{Johne1994}
A.~Johne.
\newblock Listening to the voice of the market.
\newblock {\em International Marketing Review}, 11(1):47--59, 1994.

\bibitem{Johne1998}
A.~Johne and C.~Storey.
\newblock New service development: a review of the literature and annotated
  bibliography.
\newblock {\em European Journal of Marketing}, 32(3/4):184--251, 1998.

\bibitem{Kirlappos2012}
I.~Kirlappos, M.~A. Sasse, and N.~Harvey.
\newblock Why trust seals don't work: A study of user perceptions and behavior.
\newblock In {\em Int. Conf. on Trust and Trustworthy Computing (Trust)},
  volume 7344 of {\em LNCS}, pages 308--324, Vienna, Austria, Jun 2012.
  Springer.

\bibitem{Kitchenham1996}
B.~Kitchenham and S.~L. Pfleeger.
\newblock Software quality: the elusive target.
\newblock {\em IEEE Software}, 13(1):12--21, Jan 1996.

\bibitem{Koch2017}
S.~Koch and G.~Guceri-Ucar.
\newblock Motivations of application developers: Innovation, business model
  choice, release policy, and success.
\newblock {\em J. of Organizational Computing and Electronic Commerce},
  27(3):218--238, 2017.

\bibitem{Mavridou2019}
A.~Mavridou, A.~Laszka, E.~Stachtiari, and A.~Dubey.
\newblock Verisolid: Correct-by-design smart contracts for {Ethereum}.
\newblock In {\em 23rd Int. Conf on Financial Cryptography and Data Security
  (FC)}, page to appear, St. Kitts, Feb 2019. Springer.

\bibitem{Mendelson2018}
H.~Mendelson and K.~Moon.
\newblock Modeling success and engagement for the app economy.
\newblock In {\em World Wide Web Conference (WWW)}, pages 569--578, Lyon,
  France, Apr 2018. ACM.

\bibitem{Menor2002}
L.~J. Menor, M.~V. Tatikond, and S.~E. Sampson.
\newblock New service development: areas for exploitation and exploration.
\newblock {\em J. of Operations Management}, 20(2):135--157, Apr 2002.

\bibitem{Pinna2019}
A.~Pinna, S.~Ibba, G.~Baralla, R.~Tonelli, and M.~Marchesi.
\newblock A massive analysis of {Ethereum} smart contracts empirical study and
  code metrics.
\newblock {\em IEEE Access}, 7:78194--78213, Jun 2019.

\bibitem{Porru2017}
S.~Porru, A.~Pinna, M.~Marchesi, and R.~Tonelli.
\newblock Blockchain-oriented software engineering: Challenges and new
  directions.
\newblock In {\em 39th Int. Conf. on Software Engineering Companion (ICSE-C)},
  pages 169--171, Buenos Aires, Argentina, May 2017. IEEE.

\bibitem{siegel2016understanding}
D.~Siegel.
\newblock Understanding the {DAO} attack.
\newblock {\em CoinDesk (2016)}, 2016.

\bibitem{Sigg2019}
S.~Sigg, E.~Lagerspetz, E.~Peltonen, P.~Nurmi, and S.~Tarkoma.
\newblock Exploiting usage to predict instantaneous app popularity: Trend
  filters and retention rates.
\newblock {\em ACM Trans. Web}, 13(2):13:1--13:25, Apr 2019.

\bibitem{Tonelli2018}
R.~Tonelli, G.~Destefanis, M.~Marchesi, and M.~Ortu.
\newblock Smart contracts software metrics: a first study.
\newblock Technical report, University of Cagliari, Italy, Feb 2018.

\bibitem{Venkatesh2000}
V.~Venkatesh and F.~D. Davis.
\newblock A theoretical extension of the technology acceptance model: Four
  longitudinal field studies.
\newblock {\em Management Science}, 46(2):186--204, Feb 2000.

\bibitem{Wuest2018}
K.~W\"ust and A.~Gervais.
\newblock Do you need a blockchain?
\newblock In {\em 2018 Crypto Valley Conference on Blockchain Technology
  (CVCBT)}, pages 45--54, Zug, Switzerland, Jun 2018. IEEE.

\end{thebibliography}

\end{document}